 \documentclass[one column, draft,a4paper, 11pt]{IEEEtran}

\usepackage{amsmath}
\usepackage{amsfonts}
\usepackage{amssymb}
\usepackage{amsbsy}
\usepackage{graphicx}
\usepackage{times}
\usepackage{default}
\usepackage{color}
\usepackage[noadjust]{cite}
\usepackage{enumerate}
\usepackage{tikz}
\usepackage{subcaption}

\usepackage{enumitem}

\providecommand{\lk}{\langle}
\providecommand{\rk}{\rangle}

\newcommand{\snr}{\mathsf{snr}}
\newcommand{\dof}{\mathsf{dof}}
\newcommand{\Dof}{\mathsf{DoF}}
\newcommand{\udim}{\overline{d}}
\newcommand{\ldim}{\underline{d}}

\renewcommand{\tilde}{\widetilde}
\renewenvironment{pmatrix}{\lefto (\begin{matrix}}{\end{matrix}\right )}

\def\ba#1\ea{\begin{align*}#1\end{align*}}	
\def\ban#1\ean{\begin{align}#1\end{align}}	
\def\bac#1\eac{\vspace{\abovedisplayskip}{\par\centering$\begin{aligned}#1\end{aligned}$\par}\addvspace{\belowdisplayskip}}	

\newcommand{\lefto}{\mathopen{}\left}

\newtheorem{theorem}{Theorem}

\newtheorem{definition}{Definition}
\newtheorem{proposition}{Proposition}

\newtheorem{remark}{Remark}

\IEEEoverridecommandlockouts 

\title{Characterizing Degrees of Freedom through\\[-.2cm] Additive Combinatorics}

\author{
\IEEEauthorblockN{David Stotz and  Helmut B\"olcskei\\}\vspace{.1cm}
\IEEEauthorblockA{
       Dept.~IT \& EE, ETH Zurich, Switzerland\\[-.1cm]
         {Email: \{dstotz,\ boelcskei\}@nari.ee.ethz.ch} 
\thanks{The material in this paper was presented in part at the IEEE International Symposium on Information Theory, Honolulu, HI, June~2014 \cite{SB14ISIT}.}
\thanks{The authors would like to thank M.~Einsiedler, ETH Zurich, for helpful discussions and for drawing their attention to \cite{Hoc12}.}
}}

\begin{document}
\maketitle
\begin{abstract}
We establish a formal connection between the problem of characterizing degrees of freedom (DoF) in constant single-antenna interference channels (ICs), with general channel matrix, and the field of additive combinatorics. 
The theory we develop is based on   a recent breakthrough result by Hochman in fractal geometry \cite{Hoc12}.  
Our first main contribution is an explicit condition on the channel matrix to admit full, i.e., $K/2$ DoF; this condition is satisfied for almost all channel matrices. We also provide a construction of  corresponding  DoF-optimal input distributions. 
The second main result is a  new DoF-formula exclusively in terms of Shannon entropies. This formula is more amenable to both analytical statements and numerical evaluations than the DoF-formula by Wu et al.\ \cite{WSV13}, which is in terms of R\'enyi information dimension. We then use the new DoF-formula to
shed light on the hardness of finding the exact number of DoF in ICs with rational channel coefficients, and to improve  the best known bounds on the DoF of a well-studied channel matrix.
\end{abstract}

\section{Introduction}
A breakthrough finding in network information theory was the result that $K/2$ degrees of freedom (DoF) can be achieved in $K$-user single-antenna interference channels (ICs)  \cite{CJ08,Jaf11}. The  corresponding transmit/receive scheme, known as interference alignment, exploits time-frequency selectivity of the channel  to align interference at the  receivers into low-dimensional subspaces.

Characterizing the DoF in ICs under various assumptions on the channel matrix has since  become a heavily researched topic. 
A particularly surprising result states that $K/2$ DoF can be achieved in single-antenna $K$-user ICs with constant channel matrix \cite{MGMK09, EO09}, i.e., in channels that do not exhibit any selectivity. This result was shown to hold for (Lebesgue)
almost all\footnote{Throughout the paper ``almost all'' is to be understood with respect to Lebesgue measure and ``almost sure'' is with respect to a  probability distribution that is absolutely continuous with respect to Lebesgue measure.} channel matrices \cite[Thm.~1]{MGMK09}. Instead of exploiting channel selectivity, here interference alignment happens on a number-theoretic level.
The  technical arguments---from Diophantine approximation theory---used in the proof of \cite[Thm.~1]{MGMK09} do not seem to allow an explicit characterization of the  ``almost-all set'' of full-DoF admitting channel matrices.
 What is known, though, is that channel matrices with all entries rational admit strictly less than $K/2$ DoF \cite{EO09} and hence belong to the set of exceptions relative to the ``almost-all result'' in \cite{MGMK09}.

Recently, Wu et al. \cite{WSV13} developed a general framework, based on (R\'enyi) information dimension, for characterizing the DoF in constant single-antenna ICs. While this general and elegant theory allows to recover, inter alia, the ``almost-all result'' from  \cite{MGMK09}, it does not provide insights into the structure of the set of channel matrices admitting $K/2$ DoF. In addition, the DoF-formula in \cite{WSV13} is in terms of information dimension, which can be difficult to evaluate.

\paragraph*{Contributions}
Our first main contribution  is to complement the results in \cite{MGMK09, EO09, WSV13} by providing \emph{explicit and almost surely satisfied} conditions on the IC matrix to admit full, i.e., $K/2$ DoF. The conditions we find essentially require that the set of all monomial\footnote{A monomial in the variables $x_1,...,x_n$ is an expression of the form $x_1^{k_1}x_2^{k_2}\cdots x_n^{k_n}$, with $k_i\in \mathbb N$.}
expressions in the channel coefficients  be linearly independent over the rational numbers. The proof of this result is based on  a recent breakthrough in fractal geometry  \cite{Hoc12}, which  allows us to compute the information dimension of self-similar distributions under conditions  much milder than the  open set condition \cite{BHR05} required in \cite{WSV13}. For channel matrices satisfying our explicit and almost sure conditions, we furthermore present an explicit construction of DoF-optimal input distributions. The basic idea underlying this construction has  roots in the field of additive combinatorics \cite{TV06} and essentially ensures that the set-sum of signal and interference exhibits extremal cardinality properties.
We also show that our sufficient conditions for  $K/2$ DoF are not necessary. This is accomplished by constructing examples of channel matrices that admit $K/2$ DoF but do not satisfy the sufficient conditions we identify. The set of all such channel matrices, however, necessarily has Lebesgue measure zero.  

Etkin and Ordentlich \cite{EO09} discovered that tools from additive combinatorics can be applied to characterize DoF in ICs where the off-diagonal  entries in the channel matrix are rational numbers and the diagonal entries are either irrational algebraic\footnote{A real number is called algebraic if it is the zero of a polynomial with integer coefficients. In particular, all rational numbers are algebraic.}  or rational numbers. Our second main contribution is to  establish a formal connection between additive combinatorics and the characterization of DoF in ICs with \emph{arbitrary} channel matrices. Specifically, we show
how the DoF-characterization in terms of information dimension, discovered  in \cite{WSV13}, can be translated, again based on \cite{Hoc12}, into an alternative characterization exclusively involving Shannon entropies. 
The resulting new DoF-formula  is more amenable to both analytical statements and numerical evaluation than the one in \cite{WSV13}. 
To support this statement, we show how the alternative DoF-formula can be used  to
 explain why determining the exact number of DoF for channel matrices with rational entries, even for simple examples, has remained elusive so far. Specifically, we establish that DoF-characterization for rational channel matrices is  equivalent to  very hard open problems in additive combinatorics.
Finally, we exemplify the quantitative applicability of the new DoF-formula by improving the best-known bounds on the DoF of a particular channel matrix studied in \cite{WSV13}.

\paragraph*{Notation}
Random variables are represented by uppercase letters from the end of the alphabet. Lowercase letters are used  exclusively for deterministic quantities. Boldface uppercase letters  indicate matrices.  Sets are denoted by uppercase calligraphic letters. For $x\in\mathbb R$, we write $\lfloor x \rfloor$ for the largest integer not exceeding $x$. All logarithms are taken to the  base $2$. $\mathbb E[\cdot ]$ denotes the expectation operator.  $H(\cdot)$ stands for entropy and $h(\cdot)$  for differential entropy.  For a measurable real-valued function $f$ and a measure\footnote{Throughout the paper, the terms ``measurable'' and ``measure'' are to be understood with respect to the Borel $\sigma$-algebra.} $\mu$ on its domain, the push-forward of $\mu$ by $f$  is  $(f_{\!\ast}\mu)(\mathcal A)=\mu(f^{-1}(\mathcal A))$ for  Borel sets $\mathcal A$.

\paragraph*{Outline of the paper}
In Section~\ref{sec:sysmodel}, we introduce the system model for constant single-antenna ICs. Section~\ref{sec:excond} contains  our first main result, Theorem~\ref{thm:explicit}, providing explicit and almost surely satisfied  conditions on channel matrices to admit full, i.e., $K/2$ DoF. In Section~\ref{sec:prep}, we review the basic material on information dimension, self-similar distributions, and additive combinatorics  needed in the paper. Section~\ref{sec:ideas} is devoted to 
sketching the ideas underlying the proof of  Theorem~\ref{thm:explicit} in an informal fashion and to
introducing the recent result by Hochman \cite{Hoc12} that both our main results rely on. In Section~\ref{sec:proof}, we formally prove Theorem~\ref{thm:explicit}. Section~\ref{sec:nonasy} presents  a non-asymptotic version of Theorem~\ref{thm:explicit}.  In Section~\ref{sec:notnec}, we  establish that our sufficient conditions for $K/2$ DoF are not necessary. Our second main result, Theorem~\ref{thm:eq}, which provides a DoF-characterization exclusively in terms of Shannon entropies,  is presented, along with its proof, in Section~\ref{sec:dofent}. Finally, in Section~\ref{sec:bound} we discuss the formal connection between DoF and sumset theory, a branch of additive combinatorics, and we apply the new DoF-formula to channel matrices with rational entries.

\section{System model}\label{sec:sysmodel}
We consider a  single-antenna $K$-user IC with constant channel matrix $\mathbf H =(h_{ij})_{1\leqslant i,j \leqslant K}\in\mathbb R^{K\times K}$ and input-output relation \begin{align}	Y_i={\sqrt{\snr}}\sum_{j=1}^K h_{ij} X_j +  Z_i,	\quad i=1, ... ,K ,	\label{eq:channel} \end{align}
where $X_i\in\mathbb R$ is the input at the $i$-th transmitter, $Y_i\in \mathbb R$ is the output at the $i$-th receiver, and $Z_i\in \mathbb R$ is  noise of absolutely continuous distribution such that $h(Z_i)>-\infty$ and $H(\lfloor Z_i \rfloor)<\infty$. The input signals are independent across transmitters and  noise is i.i.d.\ across  users and channel uses.

The channel matrix $\mathbf H$ is assumed to be  known perfectly at all transmitters and receivers. We impose the average power constraint
\begin{align*}	\frac{1}{n}\sum_{k=1}^n \left (x_{i}^{(k)}\right )^2\leqslant 1	
\end{align*}
on codewords $\left (x_{i}^{(1)} \,  ... \;\,  x_{i}^{(n)}\right )$ of block-length $n$ transmitted by user $i=1,...,K$.
The DoF  of this channel are defined as
\begin{align}	\Dof (\mathbf H) := \limsup_{\snr\to\infty}\frac{\overline C(\mathbf H; \snr)}{\frac{1}{2}\log \snr},	\label{eq:defdof1}	\end{align}
where $\overline C(\mathbf H; \snr)$ is the sum-capacity of the  IC.

\section{Explicit and almost sure conditions for $K/2$ DoF}\label{sec:excond}

We denote the vector consisting of the off-diagonal entries of $\mathbf H$ by $\mathbf  {\check h}\in \mathbb R^{K(K-1)}$, and  let $f_1,f_2, ...$ be the monomials  in $K(K-1)$ variables, i.e., $f_i(x_1,...,x_{K(K-1)})=x_1^{d_1}\cdots x_{K(K-1)}^{d_{K(K-1)}}$, enumerated as follows:   $f_1,...,f_{\varphi(d)}$  are  the monomials of degree\footnote{The ``degree'' of a monomial is defined as the sum of all exponents of the variables involved (sometimes called the total degree).}  not larger than $d$, where  
\ba {\varphi(d)}:=\binom{K(K-1)+d}{d} .\ea

The following theorem contains the  first main result of the paper, namely conditions on  $\mathbf H$ to admit  $K/2$ DoF that are explicit and satisfied for  almost all $\mathbf H$.

\begin{theorem}\label{thm:explicit}
Suppose that the channel matrix $\mathbf H$ satisfies the following condition:
\begin{center}	For each $i=1,...,K$, the set \ba \{f_j(	\mathbf  {\check h})  \, : \, j\geqslant 1  \}  \cup \{h_{ii} f_j(	\mathbf  {\check h}) \, : \, j\geqslant 1  \}\tag{$*$}\ea is linearly independent over $\mathbb Q$.	\end{center}
Then, we have \ba \Dof(\mathbf H) = K/2 . \ea
\end{theorem}
\begin{IEEEproof} 	See Section~\ref{sec:proof}.	\end{IEEEproof}
We first note that, as detailed in the proof of Theorem~\ref{thm:explicit}, Condition~($*$)  implies that all entries of $\mathbf H$ must be nonzero, i.e., $\mathbf H$ must be fully connected in the terminology of \cite{EO09}. By \cite[Prop.~1]{HMN05} we have $\Dof(\mathbf H)\leqslant K/2$ for fully connected channel matrices. The proof of Theorem~\ref{thm:explicit} is constructive in the sense of providing input distributions that achieve this upper bound.

Let us next dissect  Condition~($*$).
A set $\mathcal S\subseteq \mathbb R$ is linearly independent over $\mathbb Q$ if, for all $n\in \mathbb N$ and all pairwise distinct $v_1,...,v_n\in \mathcal S$, the only solution $q_1,...,q_n\in \mathbb Q$ of the equation
\ban q_1 v_1 + \ldots +q_n v_n =0 \label{eq:lincomb} \ean
is $q_1=\ldots =q_n =0$.
Thus, if Condition~($*$) is not satisfied, there exists, for at least one $i\in \{1,...,K\}$, a non-trivial linear combination of a finite number of elements of the set 
\ba \{f_j(	\mathbf  {\check h}) \, : \, j\geqslant 1  \}  \cup \{h_{ii} f_j(	\mathbf  {\check h}) \, : \, j\geqslant 1  \}\ea
with rational coefficients which equals zero. 
 In fact, this is equivalent to the existence of a non-trivial linear combination that equals zero and has all coefficients in $\mathbb Z$. This can be  seen by simply multiplying \eqref{eq:lincomb} by a common denominator of $q_1,...,q_n$.

To show that Condition~($*$) is satisfied for  almost all channel matrices, we will argue that the condition is violated on a set of Lebesgue measure zero with respect to  $\mathbf H$. To this end,
we first note that for  fixed $d\in \mathbb N$, fixed $a_1,...,a_{\varphi(d)}, b_1, ... , b_{\varphi(d)}\in \mathbb Z$ not all equal to zero, and fixed $i\in \{1,...,K\}$, 
\ban 	\sum_{j=1}^{\varphi(d)} a_jf_j(\mathbf  {\check h}) +\sum_{j=1}^{\varphi(d)} b_jh_{ii}f_j(\mathbf  {\check h}) =0 \label{eq:countable}\ean
is satisfied only on a set of measure zero with respect to $\mathbf H$, as the solutions of \eqref{eq:countable} are given by the set of  zeros of a polynomial in the  channel coefficients.
Since the set of equations \eqref{eq:countable} is countable with respect to  $d\in \mathbb N$, $a_1,...,a_{\varphi(d)}, b_1, ... , b_{\varphi(d)}\in \mathbb Z$, and $i\in\{1,...,K\}$, the set of channel matrices violating Condition~($*$)  is given by
a countable union of sets of  measure zero, which again 
has measure zero. It therefore follows that 
 Condition~($*$) is satisfied for   almost all channel matrices $\mathbf H$ and hence
Theorem~\ref{thm:explicit} provides conditions on $\mathbf H$ that not only guarantee that $K/2$ DoF can be achieved but are also explicit and almost surely satisfied.

We finally note  that the prominent example from \cite{EO09} with all entries of $\mathbf H$ rational, shown in \cite{EO09} to admit strictly less than $K/2$ DoF, does not satisfy Condition~($*$), as two rational numbers are always linearly dependent over $\mathbb Q$.

\section{Preparatory Material}
\label{sec:prep}

This section briefly reviews basic material on information dimension, self-similar distributions, and additive combinatorics needed in the rest of the paper.

\subsection{Information dimension and DoF}\label{sec:infdof}

\begin{definition}\label{def:infdim} Let $X$ be a random variable with arbitrary distribution\footnote{We consider general distributions which may be discrete,  continuous, singular, or  mixtures thereof.}  $\mu$. We define the lower and upper information dimension of $X$ as
\begin{align*} \ldim(X) := \liminf_{k\to\infty}\frac{H(\lk X \rk _k)}{\log k} \quad \text{and} \quad \udim(X):= \limsup_{k\to\infty}\frac{H(\lk X \rk _k)}{\log k},  \end{align*}
where $\lk 
X \rk_k := \lfloor kX\rfloor /k$. If  $\ldim(X) = \udim(X)$, we set $d(X):= \underline d(X) = \overline d(X)$ and  call $d(X)$ the information dimension of $X$. Since $\ldim(X), \udim(X),$ and $d(X)$ depend on $\mu$ only, we sometimes also write $\ldim(\mu), \udim(\mu),$ and $d(\mu)$, respectively. 
\end{definition}

The relevance of information dimension in characterizing DoF stems from the following relation \cite{GS07}, \cite{WSV13}, \cite{SB14} 
\ban \limsup_{\snr\to\infty}\frac{h(\sqrt{\snr} X+Z)}{\frac{1}{2}\log\snr}=\udim(X), \label{eq:guio} \ean
which holds for arbitrary independent random variables $X$ and $Z$, with the distribution of $Z$  absolutely continuous and such that  $h(Z)>-\infty$ and $H(\lfloor Z\rfloor )<\infty$. 

We can apply \eqref{eq:guio} to ICs as follows. By standard random coding arguments we get that the sum-rate 
\ban I(X_1;Y_1) + \ldots +I(X_K;Y_K) \label{eq:sumrate} \ean
is achievable, where $X_1,..., X_K$ are independent input distributions with $\mathbb E[ X_i^2] \leqslant 1$, $i=1,...,K$. Using the chain rule, we obtain
\ban &I(X_i;Y_i) \label{eq:5} = h\lefto (Y_i \right ) - h\lefto (Y_i  \; \!\vert \;\! X_i\right ) \\ & \!\!\!\! = \! h\Bigg ({\sqrt{\snr}}\sum_{j=1}^K h_{ij} X_j +  Z_i \Bigg)\! - \! h\Bigg ({\sqrt{\snr}}\sum_{j\neq i}^K h_{ij} X_j +  Z_i \Bigg ) \label{eq:7}\ean
for $i=1,...,K$. Combining \eqref{eq:guio}-\eqref{eq:7}, it now follows that  \cite{WSV13}
\begin{align} 	\dof(X_1, ... ,X_K ; \mathbf H) &:=  \nonumber \\&\!\!\!\!\! \sum_{i=1}^K \left [d\Bigg (\sum_{j=1}^K h_{ij} X_j \Bigg )-	 d \Bigg( \sum_{j\neq i}^K  h_{ij} X_j \Bigg)\right ]\label{eq:defdof} \\&\leqslant \Dof(\mathbf H), \label{eq:ineq1}   	 \end{align}
for all independent $X_1,...,X_K$ with\footnote{\label{fn:1}We only need the conditions $\mathbb E[ X_i^2] <\infty$ as scaling of the inputs does not affect $\dof(X_1, ... ,X_K ; \mathbf H) $.}  $\mathbb E[X_i^2]<\infty$, $i=1,...,K$, and such that all  information dimension terms appearing in \eqref{eq:defdof} exist. 
A striking result in \cite{WSV13} shows that inputs of discrete, continuous, or mixed discrete-continuous distribution can achieve no more than $1$ DoF irrespective of $K$. For $K>2$,  input distributions achieving $K/2$ (i.e., full) DoF therefore necessarily have a singular component.

Taking the supremum in \eqref{eq:ineq1} over all admissible  $X_1,...,X_K$ yields
\ban \Dof(\mathbf H)\geqslant  \sup_{X_1,...,X_K}\sum_{i=1}^K \left [ d\lefto (\sum_{j=1}^K h_{ij} X_j \right )-	 d \lefto ( \sum_{j\neq i}^K h_{ij} X_j \right )\right ].  \label{eq:dofwsv} \ean
It was furthermore discovered in \cite{WSV13}  that  equality in \eqref{eq:dofwsv} holds  for almost all channel matrices $\mathbf H$; 
 an explicit characterization of this ``almost-all set'', however,  does not seem to be available.
 The right-hand side (RHS) of \eqref{eq:dofwsv} can be difficult to evaluate as
explicit expressions for information dimension are available only for a few classes of distributions such as mixed discrete-continuous distributions or (singular) self-similar distributions reviewed in the next section.

\subsection{Self-similar distributions and iterated function systems}
\label{sec:ifsinf}

A  class of singular distributions  with explicit expressions for their information dimension is given by self-similar distributions \cite{Hut81}. What is more, 
self-similar input distributions can be constructed 
to retain self-similarity under linear combinations, thereby allowing us to get explicit expressions for
the information dimension of the output distributions in \eqref{eq:defdof}.
For an excellent in-depth treatment of the material reviewed in this section, the interested reader is referred to \cite{Fal04}.

We proceed to the definition of self-similar distributions.
Consider a finite set $\Phi_r:=\{\varphi_{i,r}  \, : \, i=1,..., n\}$  of  affine contractions $\varphi_{i,r} \colon \mathbb R \to \mathbb R$, i.e., \ban \varphi_{i,r}(x)=rx+w_i,\label{eq:ifs}\ean where $r\in I\subseteq (0,1)$ and the $w_i$ are pairwise distinct real numbers. We furthermore set $\mathcal W:=\{w_1,... ,w_n\}$. $\Phi_r$ is called an iterated function system (IFS)  parametrized by the contraction parameter $r\in I$. 
By classical fractal geometry \cite[Ch.~9]{Fal04} every IFS has an associated unique attractor, i.e., a non-empty compact set  $\mathcal A\subseteq \mathbb R$ such that \ban \mathcal A=\bigcup_{i=1}^n \varphi_{i,r}(\mathcal A) . \label{eq:attractor}\ean 
Moreover, for each probability vector $(p_1,...,p_n)$, there is a unique (Borel) probability distribution $\mu_r$ on $\mathbb R$ 
such that 
\ban \mu_r= \sum_{i=1}^n p_i (\varphi_{i,r})_\ast\mu_r , \label{eq:dist}\ean
where $(\varphi_{i,r})_\ast\mu_r$ is the push-forward of $\mu_r$ by $\varphi_{i,r}$. The distribution $\mu_r$ is supported on the attractor set $\mathcal A$ in \eqref{eq:attractor} and is referred to as the self-similar distribution corresponding to the IFS $\Phi_r$ with underlying probability vector $(p_1,...,p_n)$. We can give the following explicit expression for a random variable $X$ with distribution $\mu_r$ as in \eqref{eq:dist}
\ban X = \sum_{k=0}^\infty r^k W_k,		\label{eq:representation} \ean
where $\{W_k\}_{k\geqslant 0}$ is a set of  i.i.d.\ copies of a random variable $W$ drawn from the set $\mathcal W$ according to  $(p_1,...,p_n)$.

\subsection{A glimpse of additive combinatorics}
\label{ssec:addcomb}

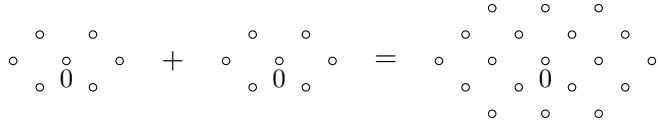
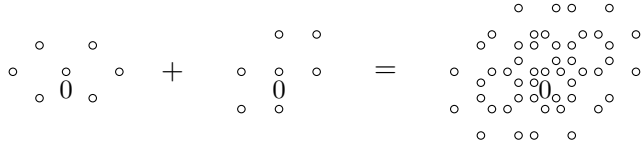
\begin{figure}
\begin{subfigure}{1\textwidth}
\begin{center}
\begin{tikzpicture}[scale=.7]
\begin{scope}
\draw (0,0) circle (2pt);
\node at (0,-.33) {\small $0$};
\draw (1,0) circle (2pt);
\draw (.5,.5) circle (2pt);
\draw (.5,-.5) circle (2pt);
\draw (-.5,.5) circle (2pt);
\draw (-.5,-.5) circle (2pt);
\draw (-1,0) circle (2pt);
\end{scope}
\node at (2,0) {$+$};
\begin{scope}[xshift=4cm]
\draw (0,0) circle (2pt);
\node at (0,-.33) {\small $0$};
\draw (1,0) circle (2pt);
\draw (.5,.5) circle (2pt);
\draw (.5,-.5) circle (2pt);
\draw (-.5,.5) circle (2pt);
\draw (-.5,-.5) circle (2pt);
\draw (-1,0) circle (2pt);
\end{scope}
\node at (6,0) {$=$};
\begin{scope}[xshift=9cm]
\draw (0,0) circle (2pt);
\node at (0,-.33) {\small $0$};
\draw (1,0) circle (2pt);
\draw (.5,.5) circle (2pt);
\draw (.5,-.5) circle (2pt);
\draw (-.5,.5) circle (2pt);
\draw (-.5,-.5) circle (2pt);
\draw (-1,0) circle (2pt);
\draw (2,0) circle (2pt);
\draw (1.5,-.5) circle (2pt);
\draw (1,-1) circle (2pt);
\draw (0,-1) circle (2pt);
\draw (-1,-1) circle (2pt);
\draw (-1.5,-.5) circle (2pt);
\draw (-2,0) circle (2pt);
\draw (-1.5,.5) circle (2pt);
\draw (-1,1) circle (2pt);
\draw (0,1) circle (2pt);
\draw (1,1) circle (2pt);
\draw (1.5,.5) circle (2pt);
\end{scope}
\end{tikzpicture}
\end{center}

\caption{Sum of two sets with common algebraic structure.}
\label{fig:bla1}

\end{subfigure}
\vspace{.7cm}

\begin{subfigure}{1\textwidth}
\begin{center}
\begin{tikzpicture}[scale=.7]
\begin{scope}
\draw (0,0) circle (2pt);
\node at (0,-.33) {\small $0$};
\draw (1,0) circle (2pt);
\draw (.5,.5) circle (2pt);
\draw (.5,-.5) circle (2pt);
\draw (-.5,.5) circle (2pt);
\draw (-.5,-.5) circle (2pt);
\draw (-1,0) circle (2pt);
\end{scope}
\node at (2,0) {$+$};
\begin{scope}[xshift=4cm]
\draw (0,0) circle (2pt);
\node at (0,-.33) {\small  $0$};
\draw (.707,.707) circle (2pt);
\draw (0,.707) circle (2pt);
\draw (.707,0) circle (2pt);
\draw (-.707,-.707) circle (2pt);
\draw (-.707,0) circle (2pt);
\draw (0,-.707) circle (2pt);
\end{scope}
\node at (6,0) {$=$};
\begin{scope}[xshift=9cm]
\draw (0,0) circle (2pt);
\node at (0,-.33) {\small  $0$};
\draw (.707,.707) circle (2pt);
\draw (0,.707) circle (2pt);
\draw (.707,0) circle (2pt);
\draw (-.707,-.707) circle (2pt);
\draw (-.707,0) circle (2pt);
\draw (0,-.707) circle (2pt);

\draw (1,0) circle (2pt);
\draw (1.707,.707) circle (2pt);
\draw (1,.707) circle (2pt);
\draw (1.707,0) circle (2pt);
\draw (.293,-.707) circle (2pt);
\draw (.293,0) circle (2pt);
\draw (1,-.707) circle (2pt);

\draw (.5,.5) circle (2pt);
\draw (1.207,1.207) circle (2pt);
\draw (.5,1.207) circle (2pt);
\draw (1.207,.5) circle (2pt);
\draw (-.207,-.207) circle (2pt);
\draw (-.207,.5) circle (2pt);
\draw (.5,-.207) circle (2pt);

\draw (.5,-.5) circle (2pt);
\draw (1.207,.207) circle (2pt);
\draw (.5,.207) circle (2pt);
\draw (1.207,-.5) circle (2pt);
\draw (-.207,-1.207) circle (2pt);
\draw (-.207,-.5) circle (2pt);
\draw (.5,-1.207) circle (2pt);

\draw (-.5,.5) circle (2pt);
\draw (.207,1.207) circle (2pt);
\draw (-.5,1.207) circle (2pt);
\draw (.207,.5) circle (2pt);
\draw (-1.207,-.207) circle (2pt);
\draw (-1.207,.5) circle (2pt);
\draw (-.5,-.207) circle (2pt);

\draw (-.5,-.5) circle (2pt);
\draw (.207,.207) circle (2pt);
\draw (-.5,.207) circle (2pt);
\draw (.207,-.5) circle (2pt);
\draw (-1.207,-1.207) circle (2pt);
\draw (-1.207,-.5) circle (2pt);
\draw (-.5,-1.207) circle (2pt);

\draw (-1,0) circle (2pt);
\draw (-.207,.707) circle (2pt);
\draw (-1,.707) circle (2pt);
\draw (-.207,0) circle (2pt);
\draw (-1.707,-.707) circle (2pt);
\draw (-1.707,0) circle (2pt);
\draw (-1,-.707) circle (2pt);

\end{scope}
\end{tikzpicture}
\end{center}

\caption{Sum of two sets with different algebraic structures.}
\label{fig:bla2}
\end{subfigure}
\caption{\small The cardinality of the sum in  (a) is $19$ and hence  small compared to the $7^2=49$  pairs  summed up, whereas the sum in  (b) has  cardinality $49$.}
\label{fig:bla}
\end{figure}

The common theme of our two main results is  a formal relationship between the study of DoF in  constant single-antenna ICs and the field of additive combinatorics. This connection is enabled by the recent breakthrough result in fractal geometry reported in \cite{Hoc12} and summarized in Section~\ref{sec:ideas}.
We next briefly discuss material from additive combinatorics that is relevant for our discussion. For a detailed treatment of additive combinatorics we refer the reader to  \cite{TV06}. 
 Specifically, we will be concerned with  sumset theory, which studies, for discrete sets $\mathcal U$, $\mathcal V$, the cardinality of the sumset $\mathcal U +\mathcal V =\{u+v   \, : \, u\in \mathcal U, v\in \mathcal V\}$ relative to $|\mathcal U|$ and $|\mathcal V|$. We begin by noting  the trivial bounds
\ban \max\{ |\mathcal U| , |\mathcal V|\} \leqslant |\mathcal U+\mathcal V| \leqslant |\mathcal U| \cdot  |\mathcal V| 	,	\label{eq:trivial} \ean 
for $\mathcal U$ and $\mathcal V$ finite and non-empty. One of the central ideas in sumset theory says that  the left-hand inequality in \eqref{eq:trivial} can be close to equality only if $\mathcal U$ and $\mathcal V$ have  a common algebraic structure (e.g., lattice structures), whereas the right-hand inequality in \eqref{eq:trivial} will be close to  equality only if the pairs $\mathcal U$ and $\mathcal V$ do not have a  common algebraic structure, i.e., they   are generic relative to each other. Figure~\ref{fig:bla} illustrates this statement.
Algebraic structures relevant in this context 
are arithmetic progressions, which are sets of the form $\mathcal S= \{a, a +d , a+2d,\ldots, a+(n-1)d \}$ with $a\in \mathbb Z$ and $d\in \mathbb N$. If $\mathcal U$ and $\mathcal V$ are finite non-empty subsets of $\mathbb Z$, an improvement of the lower bound in \eqref{eq:trivial} to $|\mathcal U | +|\mathcal V| -1 \leqslant |\mathcal U+ \mathcal V| $ can be obtained.  This lower bound is attained  if and only if $\mathcal U$ and $\mathcal V$ are arithmetic progressions of the same step size $d$ \cite[Prop.~5.8]{TV06}. 

An interesting connection between sumset theory and entropy inequalities was discovered in \cite{Ruz09, Tao10}. This connection revolves around the fact that many sumset inequalities have analogous versions in terms of entropy inequalities.
For example, the entropy version of the trivial bounds \eqref{eq:trivial} is
\ba \max \{ H(U) , H(V) \} \leqslant H(U+V) \leqslant H(U)+H(V), \ea
where $U$ and $V$ are independent discrete random variables. Less trivial examples are the  sumset inequalities \cite{TV06, Ruz96}
\ba 
|\mathcal U+\mathcal V| \cdot |\mathcal U|\cdot |\mathcal V|&\leqslant |\mathcal U-\mathcal V|^3 \\
|\mathcal U-\mathcal V| &\leqslant  |\mathcal U+\mathcal V|^{1/2} \cdot (|\mathcal U|\cdot |\mathcal V|)^{2/3},	\ea
for finite non-empty sets $\mathcal U, \mathcal V$, with their entropy counterparts  \cite{Ruz09, Tao10} 
\ban 	
H(U+V) +H(U) +H(V)&\leqslant 3 H(U-V) 	\label{eq:recall1}\\
H(U-V) &\leqslant \frac{1}{2} H(U+V)  +\frac{2}{3}( H(U) + H(V))	\label{eq:recall2}\ean
for independent discrete random variables $U, V$. Note that due to the logarithmic scale of entropy, products in sumset inequalities are replaced by sums  in their entropy versions.

\section{The cornerstones of the proof of Theorem~\ref{thm:explicit}}\label{sec:ideas}

In this section, we discuss the main ideas and conceptual components  underlying the proof of Theorem~\ref{thm:explicit}. 
First, we note that, as already pointed out in Section~\ref{sec:excond}, by \cite[Prop.~1]{HMN05} we have $\Dof(\mathbf H)\leqslant K/2$ for all $\mathbf H$ satisfying Condition~($*$).
To achieve this upper bound, we construct self-similar input distributions that yield $\dof(X_1,...,X_K;\mathbf H)=K/2$ for  channel matrices satisfying Condition~($*$).
Specifically, we take each input to have  a self-similar distribution with contraction parameter $r$, i.e.,  $X_i=\sum_{k=0}^\infty {r}^kW_{i,k}$,  where, for $i=1,...,K$,  $\{W_{i,k}  \, : \, k\geqslant 0\}$ are i.i.d.\ copies of a discrete random variable\footnote{Henceforth ``discrete random variable'' refers to a random variable  that  only takes  finitely many values.} $W_i$ with value set $\mathcal W_i$, possibly different across $i$. For the random variables $\sum_j h_{ij} X_j$ appearing in \eqref{eq:dofwsv} we then have 
\ban \sum_j h_{ij} X_j =\sum_{j} \sum_{k=0}^\infty r^k h_{ij}W_{j,k} =\sum_{k=0}^\infty r^k \sum_{j}h_{ij}W_{j,k} ,\label{eq:outputIC} \ean
and thus $\sum_j h_{ij} X_j$ is again  self-similar with contraction parameter $r$. The ``output-$\mathcal W$'' set, i.e., the value set of $\sum_{j}h_{ij}W_{j} $ is then given by $\sum_{j}h_{ij}\mathcal W_j$. 

Next, we discuss conditions on $X_j$ and $h_{ij}$ under which analytical expressions for the  information dimension of $\sum_j h_{ij} X_j$ can be given. For general self-similar distributions arising from iterated function systems 
classical results in  fractal geometry impose the so-called open set condition  \cite[Thm.~2]{GH89},
which requires the existence of a non-empty bounded set $\mathcal U\subseteq \mathbb R$ such that 
\ban 	&\bigcup_{i=1}^n\varphi_{i,r}(\mathcal U) \subseteq \mathcal U \label{eq:openset1}\\ \text{and } \quad \varphi_{i,r}(\mathcal U) &\cap \varphi_{j,r}(\mathcal U) =\emptyset, \quad \text{for all $i\neq j$},	\label{eq:openset2}\ean
for the $\varphi_{i,r}$ defined in \eqref{eq:ifs}. 
Wu et al.\ \cite{WSV13} ensure that the open set condition is satisfied by imposing an upper bound on the contraction parameter $r$ according to 
\ban r\leqslant \frac{\mathsf m (\mathcal W)}{\mathsf m (\mathcal W)+\mathsf M (\mathcal W)}, \label{eq:suff} \ean
where $\mathsf m (\mathcal W):=\min_{i\neq j} |w_i-w_j|$ and $\mathsf M (\mathcal W):=\max_{i,j} |w_i-w_j|$. 
The challenge here resides in making  \eqref{eq:suff}  hold for the output-$\mathcal W$ set. 
In  \cite{WSV13} this is accomplished by building the input sets $\mathcal W_i$ from
$\mathbb Z$-linear combinations (i.e., linear combinations with integer coefficients)  of monomials in the off-diagonal channel coefficients and then recognizing that results in Diophantine approximation theory can be used to show that \eqref{eq:suff} is satisfied for almost all channel matrices. Unfortunately, it does not seem to be possible to obtain  an explicit characterization of this  ``almost-all set''.
Recent groundbreaking work by Hochman \cite{Hoc12} replaces the open set condition by a much weaker condition, which instead of \eqref{eq:openset1}, \eqref{eq:openset2} only requires that the IFS must not  allow ``exact overlap'' of the images $\varphi_{i,r}(\mathcal A)$ and $\varphi_{j,r}(\mathcal A)$, for $i\neq j$, which we show in Theorem~\ref{thm:hochman} below   can be satisfied by ``wiggling'' with $r$ in an arbitrarily small neighborhood of its original value. This improvement turns out to be instrumental in  our  Theorem~\ref{thm:explicit} as it allows us to abandon the Diophantine approximation approach and thereby opens the doors to an explicit characterization of an ``almost-all set'' of full-DoF admitting channel matrices. 
Specifically, we use the following simple consequence of  \cite[Thm.~1.8]{Hoc12}.

\begin{theorem} \label{thm:hochman}
If $I\subseteq (0,1)$ is a non-empty compact interval which does not consist of a single point only, and $\mu_r$ is the self-similar distribution from \eqref{eq:dist} with  contraction parameter $r\in I$  and  probability vector $(p_1,...,p_n)$, then\footnote{The ``$1$'' in the minimum simply accounts for the fact that information dimension cannot exceed the dimension of the ambient space.} 
\ban d(\mu_r)= \min \lefto \{\frac{\sum p_i \log p_i}{\log r}, 1\right \} , \label{eq:formula} \ean
for all $r\in I\! \setminus\! E$, where $E$ is a set of Hausdorff and packing dimension zero.
\end{theorem}
\begin{IEEEproof}
For $\mathbf i\in \{1,...,n\}^k$, let $\varphi_{\mathbf i,r}:=\varphi_{i_1,r}\circ \ldots \circ \varphi_{i_k,r}$ and define 
\ba \Delta_{\mathbf i,\mathbf j}(r) := \varphi_{\mathbf i, r}(0) - \varphi_{\mathbf j, r}(0),  \ea
for  $\mathbf i,\mathbf j\in \{1,...,n\}^k$. Extend this definition to infinite sequences $\mathbf i,\mathbf j\in \{1,...,n\}^\mathbb N$ according to
\ba \Delta_{\mathbf i,\mathbf j}(r) := \lim_{k\to\infty}\Delta_{(i_1,...,i_k),(j_1,...,j_k)}(r).  \ea
Using  \eqref{eq:ifs} it follows that 
\ba \Delta_{\mathbf i,\mathbf j}(r) =\sum_{k=1}^\infty r^{k-1} (w_{i_k}-w_{j_k}). \ea
 Since  a power series can  vanish on a non-empty open set only if it is identically zero, we get that $\Delta_{\mathbf i,\mathbf j}\equiv 0$ on  $I$ if and only if $\mathbf i=\mathbf j$, as a consequence of the $w_i$ being pairwise distinct and $I$ containing a non-empty open set. This is precisely the condition of \cite[Thm.~1.8]{Hoc12} which asserts that
\eqref{eq:formula} holds for all $r\in I$ with the exception of a set
of Hausdorff and packing dimension zero, and thus completes the proof.
\end{IEEEproof}

\begin{remark}
Note that   \eqref{eq:formula} can be rewritten in terms of the entropy of the random variable $W$, defined in \eqref{eq:representation}, which takes value $w_i$ with probability $p_i$:
\ban d(\mu_r)= \min \lefto \{\frac{H(W)}{\log (1/r)}, 1\right \} .\label{eq:formula2}\ean
\end{remark}
\begin{remark}
 The concepts of Hausdorff and packing dimension have their roots in fractal geometry \cite{Fal04}. In the proofs of our main results, we will only need the following aspect: For $I$ as in Theorem~\ref{thm:hochman},
we can always find an $\tilde r\in  I\! \setminus\! E$  for which \eqref{eq:formula} holds. This can be seen as follows: 
$ I\! \setminus\! E=\emptyset$  implies that $E$ contains a non-empty open set and therefore would have Hausdorff and packing dimension $1$ \cite[Sec.~2.2]{Fal04}. 
\end{remark}
\begin{remark}
The strength of Theorem~\ref{thm:hochman} stems from \eqref{eq:formula} holding without any restrictions on the  $w_i\in\mathcal W$.
In particular, 
 the elements in the output-$\mathcal W$ set  $\sum_{j}h_{ij}\mathcal W_j$  may be arbitrarily close to each other rendering \eqref{eq:suff},  needed to satisfy the open set condition, obsolete.
\end{remark}

We next show how Theorem~\ref{thm:hochman}  allows us to derive explicit expressions for
the information dimension terms in \eqref{eq:defdof}. 
\begin{proposition}\label{prop:apply}
Let $r\in (0,1)$ and let $W_1,...,W_K$ be independent discrete random variables. Then, we have
\ban  \sum_{i=1}^K  &\left [  \min \lefto \{\frac{H\lefto (\sum_{j=1}^Kh_{ij}W_{j}\right )}{\log (1/  r)}, 1\right \} - \min \lefto \{\frac{H\lefto (\sum_{j\neq i}^K h_{ij}W_{j}\right )}{\log (1/  r)}, 1\right \}\right ] \leqslant \Dof(\mathbf H). \label{eq:apply} \ean
\end{proposition}
\begin{IEEEproof}
For $i=1,...,K$, let $\{W_{i,k} \, : \, k\geqslant 0\}$ be i.i.d.\ copies of $W_i$. We consider the self-similar inputs $X_i=\sum_{k=0}^\infty {r}^kW_{i,k}$,  for $i=1,...,K$.
Then, the signals 
\ba 		\sum_{j=1}^Kh_{ij}X_j &=\sum_{k=0}^\infty r^k \sum_{j=1}^Kh_{ij}W_{j,k} \\ \text{and} \quad \sum_{j\neq i}^Kh_{ij}X_j&=\sum_{k=0}^\infty r^k \sum_{j\neq i}^Kh_{ij}W_{j,k}\ea
also have self-similar distributions with contraction parameter $r$. Thus, by Theorem~\ref{thm:hochman}, for each $\varepsilon>0$, there exists an $\tilde r$ in the non-empty compact interval $I_\varepsilon:=[r-\varepsilon, r]$ (which does not consist of a single point only for all $\varepsilon>0$) such that 
\ban 	d\Bigg (\sum_{j=1}^Kh_{ij}X_j\Bigg )&= \min \lefto \{\frac{H\lefto (\sum_{j=1}^Kh_{ij}W_{j}\right )}{\log (1/ \tilde r)}, 1\right \}	\label{eq:a1}\\  \text{and} \quad d\Bigg (\sum_{j\neq i}^Kh_{ij}X_j\Bigg)&=\min \lefto \{\frac{H\lefto (\sum_{j\neq i}^K h_{ij}W_{j}\right )}{\log (1/ \tilde r)}, 1\right \}	.	\label{eq:a2}\ean
For $\varepsilon\to 0$ we have $\log(1/\tilde r)\to \log (1/r)$ by  continuity of $\log (\cdot )$. Thus, inserting \eqref{eq:a1} and \eqref{eq:a2} into \eqref{eq:ineq1} and letting $\varepsilon\to 0$, we get \eqref{eq:apply}
as desired.\end{IEEEproof}

 The  freedom we exploit in constructing full DoF-achieving $X_i$ lies in the choice of $W_1,...,W_K$ 
which thanks to Theorem~\ref{thm:hochman}, unlike in \cite{WSV13}, is not 
 restricted by distance constraints on the output-$\mathcal W$ set. 
For simplicity of exposition, we henceforth choose the same value set  $\mathcal W$ for each $W_i$.
We want to ensure that  the first term inside the sum \eqref{eq:defdof} equals $1$ and 
 the second term equals $1/2$, for all $i$, resulting in a total of $K/2$ DoF. 
It follows from \eqref{eq:a1}, \eqref{eq:a2} that 
 this can be accomplished by choosing the $W_i$ such that
\ban H\lefto (h_{ii} W_i+ \sum_{j\neq i}^Kh_{ij}W_j\right ) \approx 2 H\lefto (\sum_{j\neq i}^Kh_{ij}W_{j}\right ) \label{eq:doubling} \ean
followed by a suitable choice of the contraction parameter. 
Resorting to the analogy of entropy and sumset cardinalities sketched in Section~\ref{ssec:addcomb},  the doubling condition \eqref{eq:doubling} becomes
\ban 
\Bigg |h_{ii}\mathcal W+ \sum_{j\neq i}^Kh_{ij}\mathcal W \Bigg | &\approx \Bigg |\sum_{j\neq i}^Kh_{ij}\mathcal W \Bigg | ^2, \label{eq:key}\ean
which effectively says that the sum of the desired signal and the interference should be twice as ``rich'' as the interference alone.
Note that by the trivial lower bound in \eqref{eq:trivial}
\ban |h_{ii}\mathcal W  |  = |\mathcal W| \leqslant  \Bigg |\sum_{j\neq i}^Kh_{ij}\mathcal W \Bigg | \label{eq:key3}, \ean
and, by the trivial upper bound in \eqref{eq:trivial} 
\ban \Bigg |h_{ii}\mathcal W+ \sum_{j\neq i}^Kh_{ij}\mathcal W \Bigg | \leqslant |h_{ii}\mathcal W  |\cdot \Bigg |\sum_{j\neq i}^Kh_{ij}\mathcal W \Bigg |.  \label{eq:key2} \ean
The doubling condition \eqref{eq:key} can therefore be realized by constructing
 $\mathcal W$ such that the inequalities \eqref{eq:key3} and \eqref{eq:key2} are close to equality.
In particular, this means that (cf.\ Section~\ref{ssec:addcomb})
\begin{enumerate}[label=\Alph*)] 
\item  the terms in the sum $\sum_{j\neq i}^Kh_{ij}\mathcal W$ must have a common algebraic structure and \label{item:A}
\item  $h_{ii}\mathcal W$ and $\sum_{j\neq i}^Kh_{ij}\mathcal W$ must \emph{not} have a common algebraic structure. \label{item:B}
\end{enumerate}

The challenge here is to introduce algebraic structure into $\mathcal W$ so that \ref{item:A} is satisfied but at the same time to keep the algebraic structures of the sets $h_{ii}\mathcal W$ and $\sum_{j\neq i}^Kh_{ij}\mathcal W$ different enough so that  \ref{item:B}  is met.
Before describing the specific construction of $\mathcal W$, we note that the answer to the question of whether the sets $h_{ij}\mathcal W$  have  a common algebraic structure or not  depends on the channel coefficients $h_{ij}$.
As we want our construction to be universal in the sense of  \eqref{eq:key} holding  independently of the channel coefficients, a channel-independent choice of $\mathcal W$ is  out of the question.
Inspired by \cite{MGMK09}, we build $\mathcal W$ as a set of $\mathbb Z$-linear combinations of monomials (up to a certain degree $d\in \mathbb  N$) in the off-diagonal channel coefficients, i.e., the elements of $\mathcal W$ are given by  $\sum_{j=1}^{\varphi(d)} a_j f_j(\mathbf  {\check h})$, for $a_j \in\{1,..., N\} $ with $N\in \mathbb N$. This construction satisfies \ref{item:A} by inducing the same algebraic structure for $h_{ij}\mathcal W$, $j\neq i$, independently of the actual values of the channel coefficients $h_{ij}$, $j\neq i$. To see this, first note that multiplying
the elements $\sum_{j=1}^{\varphi(d)} a_j f_j(\mathbf  {\check h})$ of $\mathcal W$ by an off-diagonal channel coefficient $h_{ij}$, $j \neq i$, simply increases the degrees  of the participating $f_j(\mathbf  {\check h})$ by $1$. For $d$ sufficiently large the number of elements that do not appear both in  $h_{ij}\mathcal W$ and $\mathcal W$ is therefore small,  rendering  $h_{ij}\mathcal W$, $j\neq i$, algebraically  ``similar'' to $\mathcal W$, which we denote as $h_{ij}\mathcal W\approx \mathcal  W$.
We therefore get $\sum_{j\neq i}h_{ij}\mathcal W\approx \mathcal W+\ldots +\mathcal W$ as the sum of $K-1$ sets with shared algebraic structure and note that the elements of $\mathcal W+\ldots +\mathcal W$ are given by $\sum_{j=1}^{\varphi(d)} a_j f_j(\mathbf  {\check h})$ with  $a_j\in \{1,...,(K-1)N\}$.
Choosing $N$ to be large relative to $K$, we finally get 
$| \sum_{j\neq i}h_{ij}\mathcal W| \approx |\mathcal W | $. 
As for Condition~\ref{item:B}, we begin by noting that $h_{ii}$ does not participate in the monomials $f_j(\mathbf  {\check h})$ used to construct the elements in $\mathcal W$. This means that $\sum_{j\neq i}^Kh_{ij}\mathcal W$ consists of $\mathbb Z$-linear combinations of $f_j(	\mathbf  {\check h})$, while $h_{ii}\mathcal W$ consists of $\mathbb Z$-linear combinations of $h_{ii} f_j(	\mathbf  {\check h})$. By Condition~($*$)  the union of the sets $\{f_j(	\mathbf  {\check h}) \, : \, j\geqslant 1  \}$ and  $ \{h_{ii} f_j(	\mathbf  {\check h}) \, : \, j\geqslant 1  \}$  is  linearly independent over $\mathbb Q$, which ensures that  $h_{ii}\mathcal W$ and $\sum_{j\neq i}^Kh_{ij}\mathcal W$ do not share an algebraic structure.

\section{Proof of Theorem~\ref{thm:explicit} }
\label{sec:proof}

Since a set containing $0$ is always linearly dependent over $\mathbb Q$, Condition~($*$)  implies that all entries of $\mathbf H$ must be nonzero, i.e., $\mathbf H$ must be fully connected. It therefore follows from \cite[Prop.~1]{HMN05} that $\Dof(\mathbf H)\leqslant K/2$.

The remainder of the proof establishes the lower bound  $\Dof(\mathbf H)\geqslant K/2$ under Condition~($*$).
Let $N$ and $d$ be positive integers. We begin by setting
\ban \mathcal W_N:=  \Bigg \{ \sum_{i=1}^{\varphi(d)} a_i f_i(\mathbf  {\check h})   \, : \,  a_1, ..., a_{\varphi(d)}\in\{1,..., N\}  \Bigg \} \label{eq:in} \ean
and $r:=|\mathcal W_N|^{-2}$. Let  $W_1,...,W_K$ be i.i.d.\ uniform random variables on $\mathcal W_N$. By Proposition~\ref{prop:apply} we then have
\ban \sum_{i=1}^K  &\left [  \min \lefto \{\frac{H\lefto (\sum_{j=1}^Kh_{ij}W_{j}\right )}{2 \log |\mathcal W_N|}, 1\right \} \right .   \nonumber  \\ & \left .- \min \lefto \{\frac{H\lefto (\sum_{j\neq i}^K h_{ij}W_{j}\right )}{2 \log |\mathcal W_N|}, 1\right \}\right ] \leqslant \Dof(\mathbf H).\label{eq:3} \ean
Note that the random variable $\sum_{j\neq i} h_{ij}W_{j}$ takes value in 
\ban \Bigg\{\! \sum_{i=1}^{\varphi(d+1)}a_i f_i(\mathbf  {\check h})  \, : \,  a_1, ..., a_{\varphi(d+1)}\in\{1,..., (K-1)N\}  \Bigg \} . \label{eq:pres} \ean
By Condition~($*$) the set $\{f_j(	\mathbf  {\check h})  \, : \, j\geqslant 1  \} $ is linearly independent over $\mathbb Q$. Therefore, each element in the set \eqref{eq:pres} has exactly one representation as a $\mathbb Z$-linear combination with coefficients $a_1, ..., a_{\varphi(d+1)}\in\{1,..., (K-1)N\} $. This allows us to conclude that the cardinality of the set \eqref{eq:pres} is given by $((K-1)N)^{\varphi(d+1)}$, which implies $H\lefto (\sum_{j\neq i} h_{ij}W_{j}\right )\leqslant {\varphi(d+1)} \log((K-1)N) $. Similarly, we find that $|\mathcal W_N|=N^{\varphi(d)}$ and
thus  get
\ban \frac{H\lefto (\sum_{j\neq i}^K h_{ij}W_{j}\right )}{2 \log |\mathcal W_N|} &\leqslant \frac{{\varphi(d+1)}\log((K-1)N) }{2{\varphi(d)} \log N} \label{eq:2a}\\ & \xrightarrow{d,N\to \infty} \frac{1}{2}, \label{eq:2b} \ean 
where we used \ban \frac{\varphi(d+1)}{\varphi(d)}=\frac{K(K-1)+d+1}{d+1}\xrightarrow{d\to \infty}1. \label{eq:combi}\ean

We next show  that  Condition~($*$)
 implies that 
\ban H \Bigg (h_{ii} W_{i}+ \sum_{j\neq i}h_{ij}W_{j}	\Bigg ) = H \Bigg (h_{ii} W_{i}, \sum_{j\neq i}h_{ij}W_{j}	\Bigg ) . \label{eq:weshow}\ean
Applying the chain rule twice we find
\ban H \Bigg (h_{ii} W_{i}, \sum_{j\neq i}h_{ij}W_{j}	\Bigg )&= H \Bigg (h_{ii} W_{i}, \sum_{j\neq i}h_{ij}W_{j} , 	h_{ii} W_{i}+ \sum_{j\neq i}h_{ij}W_{j}	\Bigg )\\ &=  H \Bigg (h_{ii} W_{i}+ \sum_{j\neq i}h_{ij}W_{j}	\Bigg ) + H \Bigg (h_{ii} W_{i}, \sum_{j\neq i}h_{ij}W_{j}	 \; \!  \Bigg\vert   \; \!  h_{ii} W_{i}+ \sum_{j\neq i}h_{ij}W_{j}	\Bigg ), \ean
and therefore proving \eqref{eq:weshow} amounts to showing that
\ban H \Bigg (h_{ii} W_{i}, \sum_{j\neq i}h_{ij}W_{j}	 \; \!  \Bigg\vert   \; \!  h_{ii} W_{i}+ \sum_{j\neq i}h_{ij}W_{j}	\Bigg )	=0 \label{eq:amounts}. \ean
In order to establish \eqref{eq:amounts}, suppose that $w_1,...,w_K$ and $\tilde w_1,...,\tilde w_K$ are  realizations of $W_1,...,W_K$ such that 
\ban h_{ii} w_{i} + \sum_{j\neq i}h_{ij}w_{j} = h_{ii} \tilde w_{i} + \sum_{j\neq i}h_{ij} \tilde w_{j},   \label{eq:insert} \ean
or equivalently
\ban h_{ii} (w_{i}-\tilde w_i) + \sum_{j\neq i}h_{ij}(w_{j}-\tilde w_j) = 0 . \label{eq:linind}\ean
The first term on the left-hand side (LHS) of \eqref{eq:linind} is a $\mathbb Z$-linear combination of elements in  $\{h_{ii} f_j(	\mathbf  {\check h})  \, : \, j\geqslant 1  \}$, whereas the second term is a $\mathbb Z$-linear combination of elements in  $\{f_j(	\mathbf  {\check h}) \, : \, j\geqslant 1  \}$. 
Thanks to the linear independence of the union in Condition~($*$), it follows that the two terms in \eqref{eq:linind}  have to equal zero individually and hence  $w_i=\tilde w_i$ and  $\sum_{j\neq i}h_{ij}w_{j}=\sum_{j\neq i}h_{ij} \tilde w_{j}$. This shows that the sum $h_{ii} W_{i} + \sum_{j\neq i}h_{ij}W_{j}$ uniquely determines the terms $h_{ii} W_{i} $ and $\sum_{j\neq i}h_{ij}W_{j}$ and therefore proves \eqref{eq:amounts}.
Next, we note that 
\ban \! H\Bigg (\sum_{j=1}^Kh_{ij}W_{j}\Bigg )&= H\Bigg (h_{ii} W_{i}+ \sum_{j\neq i}^Kh_{ij}W_{j}\Bigg )\\ &=H\Bigg (h_{ii} W_{i}, \sum_{j\neq i}^Kh_{ij}W_{j}\Bigg )\\ &= 
H \lefto (h_{ii} W_{i} \right ) + H \Bigg ( \sum_{j\neq i}h_{ij}W_{j}	\label{eq:decomp}\Bigg ),\ean 
where the last equality is thanks to the independence of the $W_{j}$, $1\leqslant j \leqslant K$.
Putting the pieces together, we finally obtain
\ban  &\frac{H\lefto (\sum_{j=1}^Kh_{ij}W_{j}\right ) - H\lefto (\sum_{j\neq i}^K h_{ij}W_{j}\right ) }{2 \log |\mathcal W_N|}\\ &= \frac{H(h_{ii}W_{i})}{2{\varphi(d)}\log N}= \frac{{\varphi(d)}\log N}{2{\varphi(d)}\log N} =\frac{1}{2},\label{eq:1} \ean 
where we used the scaling invariance of entropy, the fact that $W_{i}$ is uniform on $\mathcal W$, and $|\mathcal W|=N^{\varphi(d)}$. 
This allows us to conclude that, for all $d$ and $N$, we have
\ban  \min \lefto \{\frac{H\lefto (\sum_{j=1}^Kh_{ij}W_{j}\right )}{2 \log |\mathcal W_N|}, 1\right \}  - \min \lefto \{\frac{H\lefto (\sum_{j\neq i}^K h_{ij}W_{j}\right )}{2 \log |\mathcal W_N|}, 1\right \}\geqslant 1- \frac{{\varphi(d+1)}\log((K-1)N) }{2{\varphi(d)} \log N} ,   \label{eq:either} \ean 
as either the first minimum on the LHS of \eqref{eq:either} coincides with the non-trivial term in which case by \eqref{eq:decomp} the second  minimum coincides with the non-trivial term as well,  and therefore by \eqref{eq:1} the LHS of \eqref{eq:either} equals $1/2\geqslant 1- \frac{{\varphi(d+1)}\log((K-1)N) }{2{\varphi(d)} \log N}$,  or the first minimum coincides with $1$ in which case we apply $\min \lefto \{\frac{H\lefto (\sum_{j\neq i}^K h_{ij}W_{j}\right )}{2 \log |\mathcal W_N|}, 1\right \}\leqslant \frac{H\lefto (\sum_{j\neq i}^K h_{ij}W_{j}\right )}{2 \log |\mathcal W_N|} \leqslant \frac{{\varphi(d+1)}\log((K-1)N) }{2{\varphi(d)} \log N}$, where we used \eqref{eq:2a} for the second inequality.
 As, by \eqref{eq:2b},  the RHS of \eqref{eq:either} converges to $1/2$ for $d,N\to \infty$, it follows that 
the LHS of  \eqref{eq:3} is asymptotically
 lower-bounded by 
$K/2$. This completes the proof.
\endIEEEproof

\section{Non-asymptotic statement}\label{sec:nonasy}
Given a channel matrix $\mathbf H$ verifying Condition~($*$) in theory requires checking  infinitely many equations of the form \eqref{eq:countable}. It is therefore natural to ask whether we can say anything about the DoF achievable for a given $\mathbf H$ when  \eqref{eq:countable} is known to hold only for finitely many coefficients $a_j,b_j$ and up to a finite degree $d$. 
To address this question we  consider the same input distributions as in the proof of Theorem~\ref{thm:explicit} and carefully analyze the steps in the proof that employ  Condition~($*$). Specifically, there are only two such steps, namely the argument on the uniqueness of the representation of elements in the set \eqref{eq:pres} and the argument leading to \eqref{eq:decomp}.
First, as to  uniqueness  in  \eqref{eq:pres} we need to verify that 
\ban 	\sum_{j=1}^{\varphi(d+1)} a_jf_j(\mathbf  {\check h})  \neq \sum_{j=1}^{\varphi(d+1)} \tilde a_jf_j(\mathbf  {\check h}) \label{eq:condfinite1}\ean
for all $a_j,\tilde a_j\in \{1,..., (K-1)N\} $ with $ (a_1,...,a_{\varphi(d+1)})\neq (\tilde a_1,...,\tilde a_{\varphi(d+1)})$. Note that we have to consider monomials up to degree $d+1$, as the multiplication of $W_j$ by an off-diagonal channel coefficient $h_{ij}$ increases the degrees of the involved monomials by $1$, as already formalized in \eqref{eq:pres}. Second, to get \eqref{eq:decomp}, we need to ensure that $h_{ii} W_{i} + \sum_{j\neq i}h_{ij}W_{j}$ uniquely determines  $h_{ii} W_{i} $ and $\sum_{j\neq i}h_{ij}W_{j}$,  for $i=1,...,K$,  which amounts to requiring $h_{ii} w_{i} + \sum_{j\neq i}h_{ij}w_{j} \neq  h_{ii} \tilde w_{i} + \sum_{j\neq i}h_{ij} \tilde w_{j}$ whenever $(h_{ii}w_i,\sum_{j\neq i}h_{ij}w_{j})\neq (h_{ii}\tilde  w_i,\sum_{j\neq i}h_{ij}\tilde  w_{j})$. 
Inserting  the elements in \eqref{eq:in} for $w_i,\tilde w_i$  this  condition reads
\ban 	\sum_{j=1}^{\varphi(d+1)} a_jf_j(\mathbf  {\check h}) +\sum_{j=1}^{\varphi(d)} b_jh_{ii}f_j(\mathbf  {\check h}) \neq \sum_{j=1}^{\varphi(d+1)} \tilde a_jf_j(\mathbf  {\check h}) +\sum_{j=1}^{\varphi(d)} \tilde b_jh_{ii}f_j(\mathbf  {\check h})  , \label{eq:condfinite2} \ean
for all $a_j,\tilde a_j\in \{1,..., (K-1)N\} $ and  $b_j,\tilde b_j\in \{1,..., N\} $ with \ba (a_1,...,a_{\varphi(d+1)},b_1,...,b_{\varphi(d)})\neq (\tilde a_1,...,\tilde a_{\varphi(d+1)},\tilde b_1,...,\tilde b_{\varphi(d)}) .\ea
Note that \eqref{eq:condfinite1} is a special case of \eqref{eq:condfinite2} obtained by setting $b_j=\tilde b_j$, for all $j$, in \eqref{eq:condfinite2}. 
Finally, rearranging terms
we find that \eqref{eq:condfinite2} simply says that non-trivial $\mathbb Z$-linear combinations of the elements participating in Condition~($*$) do not equal zero, which in turn is equivalent to   \eqref{eq:countable} restricted to a finite number of coefficients and  a finite degree.

Now, assuming that, for a given $\mathbf H$, \eqref{eq:condfinite2} is verified  for all $a_j,\tilde a_j, b_j, \tilde b_j$ and  fixed $d$ and $N$, we can proceed as in the proof of Theorem~\ref{thm:explicit} to get the following from \eqref{eq:either}:
\ba  &\min \lefto \{\frac{H\lefto (\sum_{j=1}^Kh_{ij}W_{j}\right )}{ \log (1/r)}, 1\right \}  - \min \lefto \{\frac{H\lefto (\sum_{j\neq i}^K h_{ij}W_{j}\right )}{\log (1/r)}, 1\right \}\\ &\geqslant  1- \frac{{\varphi(d+1)}\log((K-1)N) }{2{\varphi(d)} \log N}\\ &= 1- \frac{(K(K-1)+d+1)\log((K-1)N)}{2(d+1)\log N}. \ea
Upon  insertion into \eqref{eq:3} this yields the  DoF lower bound
\ba \frac{K}{2}  \left [2-\frac{(K(K-1)+d+1)\log((K-1)N)}{(d+1)\log N}\right ].  \ea

\section{Condition~($*$) is not necessary}\label{sec:notnec}

While Condition~($*$) is  sufficient  for $\Dof(\mathbf H)=K/2$, we next show that it is not necessary. This will be accomplished by constructing a class of example channel matrices that fail to satisfy Condition~($*$) but still admit $K/2$ DoF.  As, however, almost all channel matrices  satisfy Condition~($*$) this example class is necessarily of Lebesgue measure zero.
Specifically, we consider channel matrices that have $h_{ii} \in \mathbb R \setminus\! \mathbb Q$, $i=1,...,K$, and $h_{ij}\in \mathbb Q\! \setminus\! \{0\}$, for $i,j=1,...,K$ with  $i\neq j$. This assumption implies that all entries of $\mathbf H$ are nonzero, i.e., $\mathbf H$ is  fully connected, which, again by \cite[Prop.~1]{HMN05}, yields  $\Dof(\mathbf H)\leqslant K/2$. Moreover, as two rational numbers are linearly dependent over $\mathbb Q$, these channel matrices violate Condition~($*$). 
We next show that  nevertheless  $\Dof(\mathbf H)\geqslant K/2$ and hence $\Dof(\mathbf H)= K/2$. This will be accomplished by constructing corresponding  DoF-optimal input distributions. 

We begin by arguing that we may assume $h_{ij}\in\mathbb Z$, for $i\neq j$. 
Indeed, since $\Dof(\mathbf H)$ is invariant  to scaling of rows or columns of $\mathbf H$ by a nonzero constant \cite[Lem.~3]{SB14}, we can,  without affecting $\Dof(\mathbf H)$, multiply the channel matrix by a common denominator of the $h_{ij}$, $i\neq j$, thus rendering the off-diagonal entries  integer-valued 
while retaining  irrationality of the diagonal entries $h_{ii}$.

 Let 
\ban \mathcal W:= \{ 0,..., N-1\} , \label{eq:w}\ean 
for some $N>0$, and take $W_1,...,W_K$ to be i.i.d.\ uniformly distributed on $\mathcal W$. We set the contraction parameter to
\ban r= 2^{-2\log(2h_\text{max} KN)}, \label{eq:r} \ean where $h_\text{max}:=\max \{ |h_{ij}|   \, : \,  i\neq j\}$. 
Writing $\sum_{j=1}^Kh_{ij}W_{j}= h_{ii} \cdot W_{i} +1\cdot  \sum_{j\neq i} h_{ij}W_{j} $, where $W_i ,\sum_{j\neq i} h_{ij}W_{j}\in \mathbb Z$, and realizing that $\{h_{ii},1\}$ is linearly independent over $\mathbb Q$,
we can mimic the arguments leading to \eqref{eq:decomp} to conclude that \ban \! H\Bigg (\sum_{j=1}^Kh_{ij}W_{j}\Bigg )=
H \lefto (h_{ii} W_{i} \right ) + H \Bigg ( \sum_{j\neq i}h_{ij}W_{j}	\Bigg ) , \label{eq:mimic} \ean
for $i=1,...,K$.
In fact, it is precisely the linear independence of $\{h_{ii},1\}$ over $\mathbb Q$ that makes this example class work. 
Next, we note that
\ba \sum_{j\neq i}^K h_{ij}W_{j} \in \{ -h_\text{max} (K-1) N, ... ,0 , ... ,h_\text{max}(K-1)N\} \ea
and hence $ H\lefto (\sum_{j\neq i} h_{ij}W_{j}\right ) \leqslant \log \lefto (2h_\text{max}KN \right )$. 
Since the $W_{j}$, $ 1\leqslant j \leqslant K$, are identically distributed, we have $H(h_{ii}W_{i})=H(h_{ij}W_{j})$, for all $i,j$, and  therefore $H(h_{ii}W_{i})\leqslant H (\sum_{j\neq i} h_{ij}W_{j} )$ as a consequence of the fact that the entropy of a sum of independent random variableßs is greater than the entropy of each  participating random variable   \cite[Ex.~2.14]{CT06}. Thus \eqref{eq:mimic}  implies that
\ba H \Bigg (\sum_{j=1}^K h_{ij}W_{j}\Bigg )
\leqslant 2 H\Bigg(\sum_{j\neq i}^K h_{ij}W_{j}\Bigg )\leqslant  2\log \lefto (2h_\text{max}KN \right ). \ea
With \eqref{eq:r} we therefore obtain \ba \min \lefto \{ \frac{H\lefto (\sum_{j=1}^K h_{ij}W_{j}\right )}{\log (1/r)}, 1\right \}= \frac{H\lefto (\sum_{j=1}^K h_{ij}W_{j}\right )}{\log (1/r)}, \ea
and since
\ban H\lefto (\sum_{j\neq i}^K h_{ij}W_{j}\right ) \leqslant H\lefto (\sum_{j=1}^K h_{ij}W_{j}\right ) ,  \label{eq:verweise} \ean again by \cite[Ex.~2.14]{CT06}, we also have
\ba \min \lefto \{ \frac{H\lefto (\sum_{j\neq i}^K h_{ij}W_{j}\right )}{\log (1/r)}, 1\right \}= \frac{H\lefto (\sum_{j\neq i}^K h_{ij}W_{j}\right )}{\log (1/r)}. \ea
Applying Proposition~\ref{prop:apply} with \eqref{eq:mimic} and using $H(h_{ii}W_i)=\log N$, we finally obtain
\ban \Dof(\mathbf H) \geqslant  \frac{\sum_{i=1}^KH(h_{ii}W_{i})}{\log(1/r)}=  \frac{K\log N}{\log (1/r)}=\frac{K\log N}{2\log \lefto (2h_\text{max}KN \right )}.	\label{eq:lhs}\ean
Since \eqref{eq:lhs} holds for all $N$, in particular for $N\to \infty$, this establishes that $\Dof(\mathbf H)\geqslant K/2$ and thereby completes our argument.

Recall that in the case of channel matrices satisfying Condition~($*$) the value set $\mathcal W$ in \eqref{eq:in} is channel-dependent. Here, however, the assumption of the diagonal entries of $\mathbf H$ being irrational and the off-diagonal entries rational already induces enough algebraic structure for our arguments to work. In the case of  channel matrices satisfying  Condition~($*$) we induce an algebraic structure that is shared by all participating channel matrices through the choice of the channel-dependent set $\mathcal W$ and by enforcing Condition~($*$).
We conclude by noting that the example class studied here was investigated before in \cite[Thm.~1]{EO09} and \cite[Thm.~6]{WSV13}. In contrast to \cite{EO09, WSV13} our proof of DoF-optimality  is, however, not based on arguments from Diophantine approximation theory.

\section{DoF-characterization in terms of Shannon entropy}\label{sec:dofent}

To put our second main result, reported in this section, into context, we first note that the DoF-characterization \cite[Thm.~4]{WSV13}, see also \eqref{eq:dofwsv} and the statement thereafter, is in terms of information dimension. As already noted, information dimension is, in general, difficult to evaluate. Now, it turns out that the DoF-lower bound in Proposition~\ref{prop:apply} can be developed into a full-fledged DoF-characterization in the spirit of \cite[Thm.~4]{WSV13}, which, however, will be entirely in terms of Shannon entropies.

\begin{theorem}\label{thm:eq}  \; \emph{Achievability}: \; For all channel matrices $\mathbf H$, we have
\ban &\sup_{W_1,...,W_K} \frac{\sum_{i=1}^K \left [ H\lefto (\sum_{j=1}^K h_{ij} W_j \right )-	 H \lefto ( \sum_{j\neq i}^K h_{ij} W_j \right )\right ]}{ \max_{i=1,...,K}  H\lefto (\sum_{j=1}^K h_{ij} W_j \right )} \leqslant \Dof(\mathbf H),   \label{eq:dofentropy}\ean
where the supremum in \eqref{eq:dofentropy} is taken over 
all independent  discrete $W_1,...,W_K$ such that  the denominator in 
\eqref{eq:dofentropy} is nonzero.\footnote{This  condition only excludes the cases where  all $W_i$ that appear with nonzero channel coefficients are chosen as deterministic. In fact, such  choices yield $\dof(X_1, ... ,X_K ; \mathbf H)=0$  (irrespective of the choice of the contraction parameter $r$)
 and  are  thus not of  interest. }  \\ 
\emph{Converse}: \; We have equality in \eqref{eq:dofentropy}  for almost all $\mathbf H$ including channel matrices   with all off-diagonal entries  algebraic numbers and arbitrary diagonal entries.
\end{theorem}
\begin{IEEEproof}
We begin with the proof of the achievability statement. The idea of the proof is to apply Proposition~\ref{prop:apply} with a suitably chosen contraction parameter $r$.
Specifically, let $W_1,...,W_K$ be independent discrete random variables such that  the denominator in 
\eqref{eq:dofentropy} is nonzero, and  
 apply Proposition~\ref{prop:apply} with
\ba r:= 2^{-\max_{i=1,...,K}  H\lefto (\sum_{j=1}^K h_{ij} W_j \right ) },\ea
which ensures  that all minima in \eqref{eq:apply} coincide with the respective non-trivial terms.
Specifically, for $i=1,...,K$, we  have 
\ba \min \lefto \{ \frac{H\lefto (\sum_{j=1}^K h_{ij}W_{j}\right )}{\log (1/r)}, 1\right \}&= 
\frac{H\lefto (\sum_{j=1}^K h_{ij}W_{j}\right )}{\max_{i=1,...,K}  H\lefto (\sum_{j=1}^K h_{ij} W_j \right )}\\ 
\text{and\quad} \min \lefto \{ \frac{H\lefto (\sum_{j\neq i}^K h_{ij}W_{j}\right )}{\log (1/r)}, 1\right \}&
=\frac{H\lefto (\sum_{j\neq i}^K h_{ij}W_{j}\right )}{\max_{i=1,...,K}  H\lefto (\sum_{j=1}^K h_{ij} W_j \right )}, \ea 
where the latter follows from $H\lefto (\sum_{j=1}^K h_{ij} W_j \right )\geqslant H\lefto (\sum_{j\neq i}^K h_{ij} W_j \right )$ (cf.\ \eqref{eq:verweise}).  
Proposition~\ref{prop:apply} now yields
\ban \frac{\sum_{i=1}^K \left [ H\lefto (\sum_{j=1}^K h_{ij} W_j \right )-	 H \lefto ( \sum_{j\neq i}^K h_{ij} W_j \right )\right ]}{ \max_{i=1,...,K}  H\lefto (\sum_{j=1}^K h_{ij} W_j \right )} \leqslant \Dof(\mathbf H). \label{eq:takesup}\ean
Finally, the inequality \eqref{eq:dofentropy} is obtained by  supremization of the LHS of \eqref{eq:takesup} over all admissible $W_1,...,W_K$.

To prove the converse, 
we begin by  referring to the proof of \cite[Thm.~4]{WSV13}, where the following is shown to hold for almost all $\mathbf H$ including channel matrices $\mathbf H$  with all off-diagonal entries  algebraic numbers and arbitrary diagonal entries: 
For every $\delta>0$, there exist   independent discrete random variables $W_1,...,W_K$ and an $r\in (0,1)$ satisfying\footnote{This statement is obtained from the proof of \cite[Thm.~4]{WSV13} as follows.  The $W_i$ and $r$ here correspond to the $W_i$ and $r^n$ defined in \cite[Eq.~(146)]{WSV13} and \cite[Eq.~(147)]{WSV13}, respectively. The relation in \eqref{eq:fol} is then simply a consequence of \cite[Eq.~(153)]{WSV13} and the cardinality bound for entropy.}
\ban 	\log(1/r) \geqslant  \max_{i=1,...,K}  H\lefto (\sum_{j=1}^K h_{ij} W_j \right )	\label{eq:fol}\ean
 such that 
\ban \Dof(\mathbf H)\leqslant \delta + \frac{\sum_{i=1}^K \left [ H\lefto (\sum_{j=1}^K h_{ij} W_j \right )-	 H \lefto ( \sum_{j\neq i}^K h_{ij} W_j \right )\right ]}{ \log(1/r)}.  \label{eq:numerator} \ean
By \eqref{eq:fol} it follows that 
\ba  \frac{\sum_{i=1}^K \left [ H\lefto (\sum_{j=1}^K h_{ij} W_j \right )-	 H \lefto ( \sum_{j\neq i}^K h_{ij} W_j \right )\right ]}{ \log(1/r)}\leqslant \frac{\sum_{i=1}^K \left [ H\lefto (\sum_{j=1}^K h_{ij} W_j \right )-	 H \lefto ( \sum_{j\neq i}^K h_{ij} W_j \right )\right ]}{ \max_{i=1,...,K}  H\lefto (\sum_{j=1}^K h_{ij} W_j \right )}. \ea 
Finally, letting $\delta \to 0$ and taking the supremum over all admissible $W_1,...,W_K$, we get
\ba \Dof(\mathbf H)\leqslant  \sup_{W_1,...,W_K} \frac{\sum_{i=1}^K \left [ H\lefto (\sum_{j=1}^K h_{ij} W_j \right )-	 H \lefto ( \sum_{j\neq i}^K h_{ij} W_j \right )\right ]}{ \max_{i=1,...,K}  H\lefto (\sum_{j=1}^K h_{ij} W_j \right )}   \ea
for almost all $\mathbf H$ including channel matrices $\mathbf H$  with all off-diagonal entries  algebraic numbers and arbitrary diagonal entries. This completes the proof.
\end{IEEEproof}

\begin{remark}
In 
 the achievability part of Theorem~\ref{thm:eq}, we have actually shown that for all $\mathbf H$
\ban &\sup_{W_1,...,W_K} \frac{\sum_{i=1}^K \left [ H\lefto (\sum_{j=1}^K h_{ij} W_j \right )-	 H \lefto ( \sum_{j\neq i}^K h_{ij} W_j \right )\right ]}{ \max_{i=1,...,K}  H\lefto (\sum_{j=1}^K h_{ij} W_j \right )} \nonumber \\  &\leqslant  \sup_{X_1,...,X_K}\sum_{i=1}^K \left [ d\lefto (\sum_{j=1}^K h_{ij} X_j \right )-	 d \lefto ( \sum_{j\neq i}^K h_{ij} X_j \right )\right ],  \label{eq:essentially} \ean
 which combined with \eqref{eq:dofwsv} yields \eqref{eq:dofentropy}.  The LHS of  \eqref{eq:essentially} is   obtained by reasoning along the same lines as in the proof of Proposition~\ref{prop:apply},  namely by applying the RHS of  \eqref{eq:essentially} to self-similar $X_1,...,X_K$ with suitable contraction parameter $r$, invoking Theorem~\ref{thm:hochman}, and noting that the supremization  is then carried out over a smaller set of distributions. 
By Theorem~\ref{thm:eq} we know that  our alternative DoF-characterization is equivalent to the original DoF-characterization in \cite[Thm.~4]{WSV13}, i.e., \eqref{eq:essentially} holds with equality, for almost all $\mathbf H$  including $\mathbf H$-matrices  with all off-diagonal entries  algebraic numbers and arbitrary diagonal entries, since in all these cases  we have a converse for both DoF-characterizations. As shown in the next section, this includes cases where $\Dof(\mathbf H)< K/2$. Moreover,  the two DoF-characterizations are equivalent on the  ``almost-all set'' characterized by Condition~($*$), as in this case the LHS of  \eqref{eq:essentially} equals $K/2$ and therefore by \eqref{eq:dofwsv} and $\Dof(\mathbf H)\leqslant K/2$ \cite[Prop.~1]{HMN05}, we get that the RHS of \eqref{eq:essentially} equals $K/2$ as well.
What we do not know is whether \eqref{eq:essentially} is always satisfied with equality,
but certainly the set of  channel matrices where this is not the case is of Lebesgue measure zero.
\end{remark}

\begin{remark}
Compared to  the original  DoF-characterization \cite[Thm.~4]{WSV13} the alternative expression in  Theorem~\ref{thm:eq} exhibits two advantages. First, the supremization has to be carried out over discrete random variables only, whereas in \cite[Thm.~4]{WSV13}   the supremum is taken over general input distributions. Second,  Shannon entropy is typically much easier to evaluate than information dimension. Our alternative characterization is therefore more amenable to both analytical statements and numerical evaluations. 
This is demonstrated in the next section, where we put the new DoF-characterization to work to explain why determining the exact number of DoF for channel matrices with rational entries has remained elusive so far, even for simple examples. In addition, we will exemplify the quantitative applicability of our DoF-formula by improving upon the best-known bounds on the DoF of a particular channel matrix studied in \cite{WSV13}.
\end{remark}

\section{DoF characterization and additive combinatorics}
\label{sec:bound}

In this section, we apply our alternative DoF-characterization in Theorem~\ref{thm:eq} to establish a formal connection between the characterization of DoF for arbitrary channel matrices and sumset problems in additive combinatorics. We also show how Theorem~\ref{thm:eq} can be used to improve the best known bounds on the DoF of a particular channel matrix studied in \cite{WSV13}.

We begin by noting that according to \cite[Thm.~2]{EO09} channel matrices
 with all entries rational admit  strictly less than $K/2$ DoF, i.e.,
\ba \Dof(\mathbf H)<\frac{K}{2}. \ea
However, finding the exact number of DoF for rational $\mathbf H$, even for simple examples, turns out to be a very difficult problem.
Based on our alternative DoF-characterization \eqref{eq:dofentropy} in Theorem~\ref{thm:eq}, which here holds with equality as all entries of $\mathbf H$ are rational, we will be able to explain why this problem is so difficult. Specifically, we establish that characterizing the DoF for $\mathbf H$ with all entries rational is equivalent to solving very hard problems in sumset theory. As noted before, however,  finding the exact number of DoF is difficult only on a set of channel matrices of Lebesgue measure zero, since $\Dof(\mathbf H)=K/2$ for almost all $\mathbf H$.

The simplest non-trivial example is the $3$-user case with
\ba \mathbf H = \begin{pmatrix} h_1 & 0 & 0\\ h_2 &h_3 &0 \\ h_4&h_5&h_6 \end{pmatrix}, \ea 
where $h_1,...,h_6\in \mathbb Q\!\setminus \! \{ 0\}$. Since $\Dof(\mathbf H)$ is invariant  to scaling of rows or columns of $\mathbf H$ by a nonzero constant \cite[Lem.~3]{SB14}, we can transform this channel matrix as follows:\\[-.5cm]
\ba  \begin{pmatrix} h_1 & 0 & 0\\ h_2 &h_3 &0 \\ h_4&h_5&h_6 \end{pmatrix} \quad \longrightarrow \quad \begin{pmatrix} 1 & 0 & 0\\ h_2 &h_3 &0 \\ 1&\frac{h_5}{h_4}& \frac{h_6}{h_4}\end{pmatrix} \quad \longrightarrow \quad \begin{pmatrix} 1 & 0 & 0\\ h_2 &h_3 &0 \\ 1&\frac{h_5}{h_4}& 1\end{pmatrix}\quad \longrightarrow \quad \begin{pmatrix} 1 & 0 & 0\\ 1 &\frac{h_3h_4}{h_2h_5} &0 \\ 1&1& 1\end{pmatrix}.  \ea 
We can therefore restrict ourselves to the analysis of channel matrices of the form
\ban \mathbf H_\lambda = \begin{pmatrix} 1 & 0 & 0\\ 1 & \lambda &0 \\ 1&1&1 \end{pmatrix}, \label{eq:Hlambda}\ean 
where $\lambda \in \mathbb Q\!\setminus \! \{ 0\}$. 
This example class was studied before in \cite{EO09,WSV13}.
In particular, using the DoF-characterization in terms of information dimension \eqref{eq:dofwsv}, Wu et al.\  showed that \cite[Thm.~11]{WSV13} 
\ban \Dof(\mathbf H_\lambda) =1+\sup_{X_1,X_2} \left [ d(X_1+\lambda X_2) -d(X_1+X_2)  \right ], \label{eq:opt1} \ean
where the supremum is taken over all independent $X_1,X_2$ such that $\mathbb E[X_1^2],\mathbb E[X_2^2]<\infty$ and the appearing information dimension terms exist.  Based on \eqref{eq:opt1} one can lower-bound $\Dof(\mathbf H_\lambda) $ through concrete choices for the input distributions  $X_1$ and $X_2$. If one is interested in analytical expressions, these choices are, however, restricted to input distributions that allow analytical expressions for  the information dimension terms appearing in \eqref{eq:opt1}.
Upper bounds  on $\Dof(\mathbf H_\lambda) $ can be established  by employing general upper and lower bounds  on information dimension.
However, there is not much one can get beyond  what basic inequalities deliver.

By applying Theorem~\ref{thm:eq} to the channel matrix \eqref{eq:Hlambda}, we next develop an alternative characterization to  \eqref{eq:opt1}. The resulting expression for $\Dof(\mathbf H_\lambda)$ involves the minimization of the ratio of entropies of linear combinations of discrete random variables and is analytically and numerically more tractable than \eqref{eq:opt1}.
\begin{theorem}\label{thm:lambda}
For 
\ba \mathbf H_\lambda = \begin{pmatrix} 1 & 0 & 0\\ 1 & \lambda &0 \\ 1&1&1 \end{pmatrix}, \ea 
we have 
\ban \Dof(\mathbf H_\lambda) = 2 - \inf_{U, V} \frac{H(U+V)}{H(U+\lambda V)}, \label{eq:opt} \ean
where the infimum is taken over all independent discrete random variables $U,V$  such that\footnote{Again, this condition simply prevents the denominator in \eqref{eq:opt} from being zero. The case $H(U+\lambda V)=0$ is equivalent to $U$ and $V$  deterministic. This choice would, however, yield $\dof(X_1, ... ,X_K ; \mathbf H)\leqslant 1$ and is thus not of  interest.} $H(U+\lambda V)>0$.
\end{theorem}
\begin{IEEEproof}
As the 
off-diagonal entries of $\mathbf H_\lambda$ are all rational and  therefore algebraic numbers, we have equality in \eqref{eq:dofentropy}, which upon insertion of $\mathbf H_\lambda$ yields 
\ban \Dof(\mathbf H_\lambda )&= \sup_{U,V,W} \frac{  H \lefto ( U+\lambda V\right ) +H(U+V+W)-H(U+V) }{ \max  \lefto \{ H(U), H(U+\lambda V), H(U+V+W) \right \} },  \label{eq:rationalA} \ean
where the supremum is taken over all independent discrete random variables $U,V,W$ such that  the denominator in 
\eqref{eq:rationalA}  is nonzero. 
Now, again using   \cite[Ex.~2.14]{CT06}, we have   $H(U)\leqslant H(U+\lambda V)$, which when inserted into  \eqref{eq:rationalA} yields
\ban  \Dof(\mathbf H_\lambda )  &= \sup_{U,V,W} \frac{  H \lefto ( U+\lambda V\right ) +H(U+V+W)-H(U+V) }{ \max  \lefto \{  H(U+\lambda V), H(U+V+W) \right \} }  \label{eq:rationalB} \\
 & \leqslant 1 + \sup_{U,V,W} \frac{  H \lefto ( U+\lambda V\right ) -H(U+V) }{ \max  \lefto \{  H(U+\lambda V), H(U+V+W) \right \} } \label{eq:rationalC} \\
&\leqslant 1 + \sup_{U,V} \frac{  H \lefto ( U+\lambda V\right ) -H(U+V) }{  H(U+\lambda V)  } \label{eq:rationalD}  \\
 &= 2 - \inf_{U, V} \frac{H(U+V)}{H(U+\lambda V)},  \ean
where  we used the fact that the supremum in \eqref{eq:rationalC} is non-negative (as seen, e.g., by choosing $U$ to be non-deterministic and $V$  deterministic) and hence invoking $ \max  \lefto \{  H(U+\lambda V), H(U+V+W) \right \}\geqslant H(U+\lambda V)$ in the denominator of \eqref{eq:rationalC} yields the upper bound \eqref{eq:rationalD}.

For the converse part, let $U,V$  be independent discrete random variables such that $H(U+\lambda V)>0$.  We take $W$  to be discrete, independent of $U$ and $V$, and  to satisfy
\ban H(W)\geqslant H(U+\lambda V), \label{eq:wlog} \ean
e.g., we may simply choose $W$ to be uniformly distributed on a sufficiently large finite set.
Applying Proposition~\ref{prop:apply} with $W_1=U$, $W_2=V$, $W_3=W$, and $r:=2^{-H(U+\lambda V)}$, we obtain
\ban    \min \lefto \{\frac{H(U)}{H(U+\lambda V)}, 1\right \}   &+ \min \lefto \{\frac{H(U+\lambda V)}{H(U+\lambda V)}, 1\right \}   - \min \lefto \{\frac{H(U)}{H(U+\lambda V)}, 1\right \} \nonumber \\  &+\min \lefto \{\frac{H(U+ V+W)}{H(U+\lambda V)}, 1\right \} - \min \lefto \{\frac{H(U+ V)}{H(U+\lambda V)}, 1\right \} \leqslant  \Dof(\mathbf H_\lambda ).  \label{eq:implies}\ean
Since  $H(U+V+W) \geqslant H(W)\geqslant H(U+\lambda V)$, where the first inequality is by  \cite[Ex.~2.14]{CT06} and the second by the assumption \eqref{eq:wlog}, we get from \eqref{eq:implies} that
\ban    2   - \min \lefto \{\frac{H(U+ V)}{H(U+\lambda V)}, 1\right \} \leqslant  \Dof(\mathbf H_\lambda ).  \label{eq:implies2}\ean
We treat the cases $H(U+V)>H(U+\lambda V)$ and $H(U+V)\leqslant H(U+\lambda V)$ separately.
If $H(U+V)>H(U+\lambda V)$, then 
\ban  2- \frac{H(U+V)}{H(U+\lambda V)}<1 = 2   - \min \lefto \{\frac{H(U+ V)}{H(U+\lambda V)}, 1\right \} \leqslant  \Dof(\mathbf H_\lambda ).  \label{eq:implies3} \ean
On the other hand, if $H(U+V)\leqslant H(U+\lambda V)$,   \eqref{eq:implies2} becomes
\ban    2- \frac{H(U+V)}{H(U+\lambda V)}\leqslant  \Dof(\mathbf H_\lambda ).  \label{eq:implies4} \ean
Combining \eqref{eq:implies3} and \eqref{eq:implies4}, we finally get
\ban 2- \frac{H(U+V)}{H(U+\lambda V)}\leqslant \Dof(\mathbf H_\lambda ),  \label{eq:desired} \ean
for all independent $U,V$ such that $H(U+\lambda V)>0$.
Taking the supremum in \eqref{eq:desired} over all admissible $U$ and $V$ completes the proof.
\end{IEEEproof}

Through Theorem~\ref{thm:lambda} we reduced the DoF-characterization of $\mathbf H_\lambda$ to an optimization of the ratio of the entropies of two  linear combinations of discrete random variables. 
This optimization problem has a counterpart in additive combinatorics, namely the following sumset problem:
 find
finite sets $\mathcal U,\mathcal V\subseteq \mathbb R$ such that the relative size 
\ban \frac{ | \mathcal U + \mathcal V|}{| \mathcal U +\lambda \mathcal V|} \label{eq:ex} \ean
of the sumsets $ \mathcal U + \mathcal V$ and $ \mathcal U +\lambda \mathcal V$ is minimal. The additive combinatorics literature provides a considerable body of useful bounds on \eqref{eq:ex}  as a function of $|\mathcal U|$ and $|\mathcal V|$  \cite{Ruz96}.
A complete answer to this minimization problem does, however, not seem to be available.  Generally, finding the minimal value of sumset quantities as in  \eqref{eq:ex} or  corresponding entropic quantities, i.e., $H(U+V)/H(U+\lambda V)$ in this case,  appears to be a very hard problem, which indicates why finding the exact number of DoF of channel matrices with rational entries is so difficult.

The formal relationship between DoF characterization and sumset theory, by virtue of Theorem~\ref{thm:eq}, goes beyond $\mathbf H$ with rational entries and applies to general $\mathbf H$. The resulting linear combinations one has to deal with, however, quickly lead to very hard optimization problems.

We finally show how our alternative DoF-characterization can be put to use to improve the best known bounds on $\Dof(\mathbf H_\lambda)$ for $\lambda=-1$. Similar improvements are possible for other values of $\lambda$. 
For brevity we restrict ourselves, however, to the case $\lambda =-1$.

\begin{proposition}\label{prop:bounds}
We have
\ba 	1.13258 \leqslant \Dof(\mathbf H_{-1}) \leqslant \frac{4}{3}.	\ea
\end{proposition}
\begin{IEEEproof}
For the lower bound, we choose $U$ and $V$ to be independent and distributed according to
\ba 	\mathbb P[U=0]&=\mathbb P[V=0]= (0.08)^3\\ \mathbb P[U=1]&=\mathbb P[V=1]= (0.08)^2\\ \mathbb P[U=2]&=\mathbb P[V=2]= 0.08\\ \mathbb P[U=3]&=\mathbb P[V=3]=1-0.08-(0.08)^2- (0.08)^3. 	\ea
This choice is motivated by numerical investigations, not reported here.
It then follows from  \eqref{eq:opt} that
\ban   \Dof(\mathbf H_{-1}) \geqslant 2 -  \frac{H(U+V)}{H(U-V)} =	1.13258 . \ean
A more careful construction of  $U$ and $V$ should allow  improvements of this lower bound.

For the upper bound, let $U$ and $V$ be independent discrete random variables such that $H(U-V)>0$ as required in the infimum in \eqref{eq:opt}.  Recall the  entropy inequalities \eqref{eq:recall1} and \eqref{eq:recall2} stating that
\ban H(U-V) &\leqslant 3 H(U+V) - H(U) - H(V) \label{eq:tao}\\ 
H(U-V) &\leqslant \frac{1}{2} H(U+V)  +\frac{2}{3}( H(U) + H(V)). \label{eq:ruzsa} \ean 
Multiplying  \eqref{eq:tao} by $2/3$ and adding the result to \eqref{eq:ruzsa} yields
\ba \frac{5}{3} H(U-V) \leqslant \frac{5}{2} H(U+V), \ea 
and hence
\ban \frac{H(U+V)}{H(U-V)} \geqslant \frac{2}{3}.\label{eq:using} \ean 
Using \eqref{eq:using} in \eqref{eq:opt}, we then obtain
\ba \Dof(\mathbf H_{-1}) = 2 - \inf_{U, V} \frac{H(U+V)}{H(U- V)} \leqslant \frac{4}{3}, \ea
which completes the proof.
\end{IEEEproof}
The bounds  in Proposition~\ref{prop:bounds} improve on the best known bounds  obtained in \cite[Thm.~11]{WSV13}\footnote{The lower bound stated in \cite[Thm.~11]{WSV13} is actually $1.10$. Note, however, that in the corresponding proof  \cite[p.~273]{WSV13}, the term $H(U-V)- H(U+V)$ needs to be divided by $\log 3$, which seems to have been skipped  and when done leads to the lower bound $1.0681$ stated here.} 
as $1.0681\leqslant \Dof(\mathbf H_{-1}) \leqslant \frac{7}{5}$.

\bibliographystyle{IEEEtran}
\bibliography{IEEEabrv,refs}

\begin{thebibliography}{10}
\providecommand{\url}[1]{#1}
\csname url@samestyle\endcsname
\providecommand{\newblock}{\relax}
\providecommand{\bibinfo}[2]{#2}
\providecommand{\BIBentrySTDinterwordspacing}{\spaceskip=0pt\relax}
\providecommand{\BIBentryALTinterwordstretchfactor}{4}
\providecommand{\BIBentryALTinterwordspacing}{\spaceskip=\fontdimen2\font plus
\BIBentryALTinterwordstretchfactor\fontdimen3\font minus
  \fontdimen4\font\relax}
\providecommand{\BIBforeignlanguage}[2]{{%
\expandafter\ifx\csname l@#1\endcsname\relax
\typeout{** WARNING: IEEEtran.bst: No hyphenation pattern has been}%
\typeout{** loaded for the language `#1'. Using the pattern for}%
\typeout{** the default language instead.}%
\else
\language=\csname l@#1\endcsname
\fi
#2}}
\providecommand{\BIBdecl}{\relax}
\BIBdecl

\bibitem{SB14ISIT}
D.~Stotz and H.~B\"olcskei, ``Explicit and almost sure conditions for {$K/2$}
  degrees of freedom,'' \emph{Proc. IEEE Int. Symp. on Inf. Theory}, pp.
  471--475, June 2014.

\bibitem{Hoc12}
M.~Hochman, ``On self-similar sets with overlaps and inverse theorems for
  entropy,'' \emph{Annals of Mathematics}, Vol. 180, No.~2, pp. 773--822,
  {Sep.} 2014.

\bibitem{WSV13}
Y.~Wu, S.~Shamai~(Shitz), and S.~Verd\'u, ``A formula for the degrees of
  freedom of the interference channel,'' \emph{IEEE Trans. Inf. Theory},
  Vol.~61, No.~1, pp. 256--279, {Jan.} 2015.

\bibitem{CJ08}
V.~R. Cadambe and S.~A. Jafar, ``Interference alignment and degrees of freedom
  of the {K}-user interference channel,'' \emph{IEEE Trans. Inf. Theory},
  Vol.~54, No.~8, pp. 3425--3441, Aug. 2008.

\bibitem{Jaf11}
S.~A. Jafar, ``Interference alignment --- {A} new look at signal dimensions in
  a communication network,'' \emph{Foundations and Trends in Communications and
  Information Theory}, Vol.~7, No.~1, 2011.

\bibitem{MGMK09}
A.~S. Motahari, S.~O. Gharan, M.-A. Maddah-Ali, and A.~K. Khandani, ``Real
  interference alignment: Exploiting the potential of single antenna systems,''
  \emph{IEEE Trans. Inf. Theory}, Vol.~60, No.~8, pp. 4799--4810, {June} 2014.

\bibitem{EO09}
R.~H. Etkin and E.~Ordentlich, ``The degrees-of-freedom of the {K}-user
  {Gaussian} interference channel is discontinuous at rational channel
  coefficients,'' \emph{IEEE Trans. Inf. Theory}, Vol.~55, No.~11, pp.
  4932--4946, {Nov.} 2009.

\bibitem{BHR05}
C.~Brandt, N.~Viet~Hung, and H.~Rao, ``On the open set condition for
  self-similar fractals,'' \emph{Proc. of the AMS}, Vol. 134, No.~5, pp.
  1369--1374, {Oct.} 2005.

\bibitem{TV06}
T.~Tao and V.~Vu, \emph{Additive Combinatorics}, ser. Cambridge {S}tudies in
  {A}dvanced {M}athematics.\hskip 1em plus 0.5em minus 0.4em\relax New York,
  NY: Cambridge University Press, 2006, Vol. 105.

\bibitem{HMN05}
A.~H{\o}st-Madsen and A.~Nosratinia, ``The multiplexing gain of wireless
  networks,'' \emph{Proc. IEEE Int. Symp. on Inf. Theory}, pp. 2065--2069,
  {Sep.} 2005.

\bibitem{GS07}
A.~Guionnet and D.~Shlyakhtenko, ``On classical analogues of free entropy
  dimension,'' \emph{Journal of Functional Analysis}, Vol. 251, pp. 738--771,
  {Oct.} 2007.

\bibitem{SB14}
D.~Stotz and H.~B\"olcskei, ``Degrees of freedom in vector interference
  channels,'' \emph{Submitted to IEEE Trans. Inf. Theory, arXiv:1210.2259v2},
  Vol. cs.IT, {Sep.} 2014.

\bibitem{Hut81}
J.~E. Hutchinson, ``Fractals and self similarity,'' \emph{Indiana University
  Mathematics Journal}, Vol.~30, pp. 713--747, 1981.

\bibitem{Fal04}
K.~Falconer, \emph{Fractal Geometry: Mathematical Foundations and
  Applications}, {2nd}~ed.\hskip 1em plus 0.5em minus 0.4em\relax John Wiley \&
  Sons, 2004.

\bibitem{Ruz09}
I.~Ruzsa, ``Sumsets and entropy,'' \emph{Random Structures {\&} Algorithms},
  Vol.~34, No.~1, pp. 1--10, {Jan.} 2009.

\bibitem{Tao10}
T.~Tao, ``Sumset and inverse sumset theory for {Shannon} entropy,''
  \emph{Combinatorics, Probability {\&} Computing}, Vol.~19, No.~4, pp.
  603--639, July 2010.

\bibitem{Ruz96}
I.~Z. Ruzsa, ``Sums of finite sets,'' in \emph{Number Theory: New York Seminar
  1991--1995}, D.~V. Chudnovsky, G.~V. Chudnovsky, and M.~B. Nathanson,
  Eds.\hskip 1em plus 0.5em minus 0.4em\relax Springer US, 1996, pp. 281--293.

\bibitem{GH89}
J.~S. Geronimo and D.~P. Hardin, ``An exact formula for the measure dimensions
  associated with a class of piecewise linear maps,'' \emph{Constructive
  Approximation}, Vol.~5, pp. 89--98, {Dec.} 1989.

\bibitem{CT06}
T.~M. Cover and J.~A. Thomas, \emph{Elements of Information Theory},
  2nd~ed.\hskip 1em plus 0.5em minus 0.4em\relax New York, NY:
  Wiley-Interscience, 2006.

\end{thebibliography}
\end{document}